\documentclass[aps,prd,twocolumn,groupedaddress,showpacs]{revtex4}
\usepackage{graphicx}% Include figure files
\usepackage{dcolumn}% Align table columns on decimal point
\usepackage{bm}% bold math
\usepackage{amssymb,amsmath,latexsym,amsfonts}
\begin{document}

\title{Polymer quantization in the Bogoliubov's regime for a homogeneous one-dimensional Bose-Einstein condensate}

\author{El\'ias Castellanos}
\email{ecastellanos@mctp.mx.} \affiliation{Mesoamerican Centre for Theoretical Physics \\ (ICTP regional headquarters in Central America, the Caribbean and Mexico).\\  Universidad Aut\'onoma de Chiapas.
Ciudad Universitaria, Carretera Zapata Km. 4, Real del Bosque (Ter\'an), 29040, Tuxtla Guti\'errez, Chiapas, Mexico.}

\author{Guillermo Chac\'{o}n-Acosta}
\email{gchacon@correo.cua.uam.mx} \affiliation{Departamento de
Matem\'aticas Aplicadas y Sistemas, \\ 
Universidad Aut\'onoma Metropolitana-Cuajimalpa,\\
Vasco de Quiroga 4871,  M\'exico D. F., 05348, Mexico}

\author{H\'ector H. Hern\'andez-Hern\'andez}
\email{hhernandez@uach.mx} \affiliation{ Facultad de Ingenier\'ia\\Universidad Aut\'onoma de Chihuahua\\
Nuevo Campus Universitario, Chihuahua 31125, Mexico.}

\author{El\'i Santos}
\email{esantos@mctp.mx.} \affiliation{Mesoamerican Centre for Theoretical Physics \\ (ICTP regional headquarters in Central America, the Caribbean and Mexico).  Universidad Aut\'onoma de Chiapas.\\
Ciudad Universitaria, Carretera Zapata Km. 4, Real del Bosque (Ter\'an), 29040, Tuxtla Guti\'errez, Chiapas, Mexico.}
\begin{abstract}
In the present report we analyze the eventual modifications caused by the polymer quantization upon the ground state of a homogeneous one-dimensional Bose--Einstein condensate. We obtain the ground state energy of the corresponding N--body system and, consequently, the corresponding speed of sound, allowing us to explore the sensitivity of the system to corrections caused by the polymer quantization. The corrections arising from the polymer quantization can be improved for dense systems together with small values of the corresponding one--dimensional scattering length. However, these corrections remain constrained due to finite size effects of the system. The contributions of the polymer length scale to the properties of the ground state energy of the system allow us to explore, as a first approximation and when the Bogoliubov's formalism is valid, the sensitivity of this many--body system to traces caused by the discreteness of space suggested by the polymer quantization.
\end{abstract}

%\date{\today}
\maketitle

\section{Introduction}

One interesting feature, related to Bose--Einstein condensates, is its possible use as test tools in gravitational physics \cite{eli,CastellanosCamacho,CastellanosCamacho1,Castellanos,r1,r2,elca,eli2,CastellanosClaus,ECA,ECA1,ECe}. The quantum properties associated with this N--body system suggest that this phenomenon could be taken as a serious candidate to explore some features related to a possible quantum structure of space--time. A possible discreteness of space may lead, for instance, to anomalous dispersion relations, deformations of the Heisenberg's uncertainty principle, and non--commutative geometries that may be related to the so--called polymer quantization \cite{AS1,AS2}. Such deformations are a general feature of quantum--gravity models, for instance, Loop Quantum Gravity  \cite{Gam,hugo} or Non--commutative Geometries \cite{gio1,j,O}. Furthermore, it is also of great relevance in the general context of Lorentz--symmetry breaking \cite{sr}.

As mentioned above, some basic properties related to Bose--Einstein condensates, for instance, the condensation temperature \cite{Castellanos,CastellanosClaus}, the corresponding speed of sound \cite{eli2,ECA} and also the free velocity expansion and the interference fringes of two overlapping condensates \cite{ECA1}, have been used to explore the sensitivity of these systems to a possible manifestation of the Planck scale regime. 

In the same spirit, in a previous report \cite{eli} the polymer quantization  %\cite{AS1,AS2}
has been applied to a one--dimensional Bose-Einstein condensate, for which bounds on the so--called polymer length were obtained under typical laboratory conditions by means of the shift on the condensation temperature. Additionally, in Ref.\,\cite{chaca} polymer quantization was applied also to compact star models in order to explore the corrections in some basic properties of these systems caused by the introduction of the polymer length. However, the impossibility to directly observe  the properties of these systems from an accurate experimental point of view, makes low--energy, earth--based experiments, for instance, Bose--Einstein condensates, a more interesting and also a more relevant tool in searching for some features related to a posible quantum (discreteness) structure  of space--time.
 
Polymer quantum systems are quantum mechanical models quantized in a similar way as in loop quantum gravity, that allow the study of the discreteness of space and other features of the loop quantization in a simplified form \cite{vel}.      
In the \textit{loop or polymer}
representation the corresponding Hilbert space $\mathcal{H}_{poly}$ is
spanned by the basis states $\{|x_j\rangle\}$, whose coefficients have a suitable
fall-off \cite{AS1}, endowed with the inner product $ \langle x_i|x_j\rangle  = \delta_{ij}$ where $\delta_{i j}$ is the Kronecker delta. 

The basic ingredients of this quantization are the position and the translation operators. The position operator $\hat{x}$ acts as usual, i.e.,
$\hat{x}|x_j\rangle=x_j|x_j\rangle\,,$ 
while the translation operator $\hat{V}(\mu)$ shifts the state to an arbitrary distance position $\mu$, that is the length of the lattice, where $x_j$ takes values,
$\hat{V}(\mu)|x_j\rangle=|x_j-\mu \rangle.$
In the polymer representation the translation operator $\hat{V}(\mu)$ is not weakly continuous
with respect to  $\mu$, so there is no well-defined momentum operator as compared to the standard quantum mechanics, due to the discrete structure assigned to space \cite{AS1}.

Under these circumstances the polymer Hamiltonian can be defined as follows
\begin{equation}\label{ham}
  \widehat{H}_{\mu} = \frac{\hbar^2}{2m\mu^2} \left[ 2- \hat{V}(\mu) -\hat{V}(-\mu)\right]+ \hat{U}(x),
\end{equation}
where $\hat{U}(x)$ is the potential term, which in this manuscript we will assume as a square well potential. In other words, we analyze the case of a homogeneous  one--dimensional Bose--Einstein condensate. 

Following \cite{AS1,AS2} the Hamiltonian (\ref{ham}) can be formally written as follows
\begin{equation}\label{ham-sin}
     \widehat{H}_{\lambda}= \frac{\hbar^{2}}{2m\lambda^2}\widehat{\sin^{2}\Bigl(\frac{\lambda
p}{\hbar}\Bigr)} + \hat{U}(x),
\end{equation}
where $\lambda$ is the so-called {\it polymer length} defined by $\lambda = 2\mu$, and can be considered as a fundamental length scale.
We can use the last expression to obtain the effective polymer Hamiltonian by replacing the kinetic term with the square of the sine function.

As a toy model we will explore the properties of a homogeneous one--dimensional Bose--Einstein condensate for which the Bogoliubov's formalism is valid \cite{sp1}, corrected by the contributions of the polymer quantization \footnote{In this work we employed the effective Hamiltonian (\ref{effective-Hamiltonian}) to determine the effects of the introduction of the polymer length $\lambda$ on Bose-Einstein condensates. However, there is a more general way of doing this: one can obtain an expression for the kinetic term in (\ref{ham-sin}) in terms of creation and annihilation operators as in (\ref{effective-Hamiltonian}), see for instance\cite{viqar}, although it is much more involved.}. This will allow us to extend our results to more realistic scenarios, namely, trapped Bose--Einstein condensates, in future work. Although a one--dimensional condensate is not physically feasible, it is possible to obtain a quasi-one--dimensional condensate by using extremely anisotropic traps  (see for instance Refs \cite{Karabulut,yan,Dalfovo}, and \,\cite{BP,TK}).\\
This opens the opportunity to explore possible modifications to the properties of these systems by employing the polymer quantization. As a first attempt we analyze the energy of the ground state and the corresponding speed of sound of a homogeneous one-dimensional Bose--Einstein condensate.

\section{Polymer Bogoliubov Spaces}

Let us start with the one--dimensional effective polymer Hamiltonian for a single particle \cite{chiapas} given by
\begin{equation}
H=\frac{\hbar^{2}}{2m\lambda^2}\sin^{2}\Bigl(\frac{\lambda
p_{x}}{\hbar}\Bigr) \label{Ham0},
\end{equation}
where $\lambda$ is the so--called polymer
length scale. 

The semiclassical single particle energy spectrum
associated with (\ref{Ham0}) can be expressed as follows
\begin{equation}\label{ep}
\epsilon_{p_{\lambda}}=\frac{\hbar^{2}}{2m\lambda^2}\sin^{2}\Bigl(\frac{\lambda
p_{x}}{\hbar}\Bigr).
\end{equation}
Assuming that $\lambda<<1$, we can expand the semiclassical energy spectrum (\ref{ep}), to second order in the polymer length scale $\lambda$ 
\begin{equation}
\label{ep1}
\epsilon_{p_{\lambda}} = \frac{p^{2}}{2m}-\frac{\lambda^{2}}{2m\hbar^{2}} p^{4}+... 
\end{equation}
In order to obtain the energy of the ground state of a one--dimensional condensate, let us propose the following N--body one--dimensional  Hamiltonian corrected by the introduction of the polymer length scale $\lambda$
\begin{eqnarray}\label{effective-Hamiltonian}
\hat{H} =
\sum_{p=0} \frac{p^2}{2m}\Bigl[1-\lambda^{2} \Bigl(\frac{p}{\hbar}\Bigr)^{2} \Bigr]\,\, \hat{a}_{p}^{\dagger}\hat{a}_{p} \nonumber\\
+\frac{U_{0_{1D}}}{2L}\sum_{p=0}\sum_{q=0}\sum_{r=0}\hat{a}_{q}^{\dagger}\hat{a}_{r}^{\dagger}
 \hat{a}_{q+p}\hat{a}_{r-p} , \label{Ham1}
\end{eqnarray}
where $U_{0_{1D}}$ is the one dimensional self interaction parameter which describes the interactions within the system, with $L$ the characteristic length (width) of the system. We assume also that the creation and annihilation operators $(\hat{a}_{p}^{\dagger},\ \hat{a}_{p})$,  satisfy the usual canonical commutation relations for bosons. We want to mention that the ladder operators in the context of a polymer, background independent quantum field theory, have been defined in \cite{viqar} as a realization of Fock-like states, and reduces to those employed above when the polymer scale vanishes. Here we do not use this formalism as we seek an approximation to lowest order in $\lambda^2$.

%Additionally, the case $p=0$ depicts the corresponding ground state level.
Furthermore, we assume that below the condensation temperature the number of particles in the ground state is approximately equal to the total number of particles and the number of particles in the excited states is negligible. 
%
%\begin{equation}
%N_{0}\approx N, \,\,\,\,\,\,\,\, \sum_{p \not=0} N_{p} <<N,
%\end{equation}
%being $N$ the total number of particles, $N_{p}$ the number of particles in the excited states, and $N_{0}$ the number of particles in the ground state. 
%
It is noteworthy to mention that one--dimensional Bose--Einstein condensates have a pathological behavior in the thermodynamic limit \cite{yuka1}. In order to make the condensation possible, finite size effects of the system need to be taken into account and, consequently, the contribution to the ground state energy per particle must also be considered. In other words, we assume that the ground state is not given by $p=0$, instead we will assume that the minimum value for the momentum $p$ is given by $p_{0}=\hbar /L$ \cite{yuka1}.  Thus, we will prove that when the Bogoliubov's approximation is valid and when the polymer length is set to zero we recover the results obtained in Refs.\,\cite{paper60,paper60_1}, corrected by the finite size effects of the system as in Ref.\,\cite{sp1}. 

If only terms up to second order in $\hat{a}_{0}$ are relevant then, within the same order of the approximation, this implies that $ \langle \hat{a}_{0}^{\dag} \hat{a}_{0} \rangle \approx  \langle \hat{a}_{0}^{2} \rangle \approx  \langle \hat{a_{0}^{\dag^{2}}}  \rangle \approx  N$. 

In this scenario the Hamiltonian (\ref{Ham1}) is given by
\begin{eqnarray}
\hat{H} &= &\frac{U_{0_{1D}}N^2}{2L}+ \frac{(\hbar/L)^{2}}{2m}\Bigl[1-\Bigl(\frac{\lambda}{L}\Bigr)^{2}\Bigr]\,N \nonumber\\ &+& 
\sum_{p \not=p_0}\Bigg[\frac{p^{2}}{2m}\Bigl[1-\lambda^{2} \frac{p^{2}}{\hbar^{2}} \Bigr]+\frac{U_{0_{1D}}N}{L}\Bigg] \hat{a}_{p}^{\dagger}\hat{a}_{p} \nonumber\\ &+& \sum_{p \not=p_0}\frac{U_{0_{1D}} N}{2L}\Bigl[\hat{a}_{p}^{\dagger}\hat{a}_{-p}^{\dagger}
+ \hat{a}_{p}\hat{a}_{-p}\Bigr]. 
\label{Ham2}
\end{eqnarray}
The second term in Eq.\,(\ref{Ham2}) contains contributions due to the finite size effects of the system and also the contributions from the polymer length scale $\lambda$ as well, which in fact scales with the number of particles. 

The Hamiltonian for a one--dimensional Bose--Einstein condensate with a $\delta$--interacting potential can be diagonalized via the Bethe \emph{anzats} \cite{paper60}. The energy per particle was deduced in this scenario, giving
\begin{equation}
\label{paper60}
E_{n}=\frac{\hbar^{2}}{2m}n^{2}e[g(n)],
\end{equation}
where $e[g(n)]$ is a function of the parameter $g(n)=U_{0_{1D}}/n$, with $n$ the corresponding density of particles. 
Equation (\ref{paper60}) has the two following  limiting cases
\begin{equation}
\label{LD}
E_{n} \approx \frac{\pi^{2} \hbar^{2}}{6m} n^{2},
\end{equation} 
\begin{equation}
\label{HD}
E_{n} \approx \frac{U_{0_{1D}}}{2}n.
\end{equation}
The limiting case Eq.\,(\ref{LD}) corresponds to the low density limit associated with the case of infinitely strong interactions of a gas of impenetrable bosons or Tonks--Girardeau gas \cite{tonks}. This expression formally coincides with that for free fermions, which suggest some kind of Fermi--Bose duality in one--dimensional systems \cite{tonks,FB}.

On the other hand, Eq.\,(\ref{HD}) depicts the high density limit. This limit corresponds to the Thomas--Fermi energy functional \cite{cigarro} first introduced by Bogoliubov \cite{Bog,Bog1}. We must mention that in the three--dimensional case the mean--field approximation is valid for low densities, contrary to the one--dimensional case where it is valid at high densities. In other words, the Bogoliubov theory should be correct for small values of $g(n)$. Thus, if we assume that the Bogoliubov's formalism is valid, that is, if the system lies in the high density limit, then, without loss of generality, we may apply from the very beginning the pseudo--potential method in order to diagonalize our Hamiltonian (\ref{Ham2}). However, as we will see later, a novel ingredient is needed, namely, finite size effects are also required. In fact, in Ref.\,\cite{sp1} the corrections caused by finite size effects in this type of systems were recently analyzed. It was showed there that when the Bogoliubov's formalism is valid then finite size effects must be taken into account in order to obtain  well defined ground state properties. 

The Hamiltonian (\ref{Ham2}) can be diagonalized by introducing, as usual within this approximation, the so--called Bogoliubov's transformations \cite{pathria,Ueda} 
\begin{equation}
\hat{a}_{p}=
\frac{\hat{b}_{p} -
\alpha_p \hat{b}_{-p}^{\dagger}}{\sqrt{1-\alpha_p^2}},\label{Bog1} \hspace{0.5cm}\hat{a}_{p}^{\dagger}=
\frac{\hat{b}_{p}^{\dagger} -
\alpha_p\hat{b}_{-p}}{\sqrt{1-\alpha_p^2}},
\end{equation}
where $\alpha_{p}$ is the Bogoliubov's coefficient. The operators $\hat{b}_{p}^{\dagger}$ and $\hat{b}_{p}$ are creation and annihilation operators that also obey the canonical commutation relations for bosons \cite{pathria}. 
Inserting the Bogoliubov transformations (\ref{Bog1}) into the Hamiltonian (\ref{Ham2}), we are able to obtain the following diagonalized Hamiltonian
\begin{eqnarray}
\label{Ham3}
\hat{H} &=& \frac{U_{0_{1D}}N^2}{2L}+ \frac{(\hbar/L)^{2}}{2m}\Bigl[1-\Bigl(\frac{\lambda}{L}\Bigr)^{2}\Bigr]\,N + \nonumber\\ && \sum_{p \not=0}\sqrt{\epsilon_{p_{\lambda}}\Bigl(\epsilon_{p_{\lambda}}+\frac{2U_{0_{1D}}N}{L}\Bigl)}\,\,
\hat{b}_{p}^{\dagger}\hat{b}_{p} +\nonumber\\&&\sum_{p \not=0}\Bigg\{-\frac{1}{2}\Bigg[\frac{U_{0_{1D}}N}{L} +\epsilon_{p_{\lambda}}
 \nonumber\\ &-&\sqrt{\epsilon_{p_{\lambda}}\Bigl(\epsilon_{p_{\lambda}}+\frac{2U_{0_{1D}}N}{L}\Bigr)}\Bigg]\Bigg\}. \,\,\,\,\,\,\,\,\,\,\,\,
\end{eqnarray}
%
%In the usual case, i.e., without polymer length contributions, the last summation in the Hamiltonian (\ref{Ham3}) would be divergent as $(U_{0_{1D}}N/L)^{2}/2\epsilon_{p}$, as can be seen by performing an expansion of the last term in equation (\ref{Ham3}) for large $p$. 

We also assume that the pseudo--potential $U_{ps}(x)$ can be expressed as $ U_{ps}(x)=U_{0_{1D}}\delta(x)d/d\, x(x *\_)$ where $\delta(x)$ is the Dirac delta function.  
Notice that in our case, the Hamiltonian (\ref{Ham3}) diverges as $(1/p^{2})+2m\lambda^{2}/\hbar^{2}$ for large $p$ and $\lambda^2 << 1$. Thus, for all practical purposes, the divergence can be expressed as $(1/p ^{2})+cte$. This yields to $1/x+\delta(x)$ in the configuration space after the application of the Fourier transform. Fortunately, and due to the property $x\delta(x)=0$, the constant term does not contribute to the pseudo potential $U_{ps}(x)$. These facts reinforce our approach, i.e., when the Bogoliubov's formalism is valid, it could be useful to describe some properties of our \emph{polymer} condensate. In other words, even in this scenario the action of the pseudo potential $U_{ps}(x)$ removes the divergence. 

Thus, by using the pseudo--potential method, our Hamiltonian (\ref{Ham3}) now is well defined and can be re--expressed as follows
\begin{eqnarray}
\label{Ham301}
\hat{H} &=& \frac{U_{0_{1D}}N^2}{2L} + \frac{(\hbar/L)^{2}}{2m}\Bigl[1-\Bigl(\frac{\lambda}{L}\Bigr)^{2}\Bigr]\,N + \nonumber\\ && \sum_{p \not=p_0}\sqrt{\epsilon_{p_{\lambda}}\Bigl(\epsilon_{p_{\lambda}}+\frac{2U_{0_{1D}}N}{L}\Bigl)}\,\,
\hat{b}_{p}^{\dagger}\hat{b}_{p} +\nonumber\\&&\sum_{p \not=p_0}\Bigg\{-\frac{1}{2}\Bigg[\frac{U_{0_{1D}}N}{L} +\epsilon_{p_{\lambda}}
\\\nonumber&-&\sqrt{\epsilon_{p_{\lambda}}\Bigl(\epsilon_{p_{\lambda}}+\frac{2U_{0_{1D}}N}{L}\Bigr)}-\Bigl(\frac{U_{0_{1D}}N}{L}\Bigr)^2\frac{1}{2 \epsilon_{p_{\lambda}}}\Bigg]\Bigg\}. 
\end{eqnarray}

In order to obtain the ground state energy of the system we replace the last summation in the Hamiltonian (\ref{Ham301}) by an integration, as usual. Then, the following expression associated with the ground state energy of the system when $\lambda^{2}<<1$ is obtained   
\begin{eqnarray}
E_{0}&\approx& \frac{U_{0_{1D}}N^2}{2L} + \frac{(\hbar/L)^{2}}{2m}\Bigl[1-\Bigl(\frac{\lambda}{L}\Bigr)^{2}\Bigr]\,N \nonumber\\ &-&\frac{L\sqrt{2m}}{4\pi\hbar}\Bigl(\frac{U_{0_{1D}}N}{L}\Bigr)^{3/2}\int_{\gamma}^{\infty} f(z)\,dz \nonumber\\&+&
3\lambda^{2} \frac{L\sqrt{2m^{3}}}{8\pi\hbar^{3}}\Bigl(\frac{U_{0_{1D}}N}{L}\Bigr)^{5/2} \int_{\gamma}^{\infty}z^{2}\,f(z)\,dz,
\label{Psedo2}
\end{eqnarray}
where we have defined the dimensionless variable $z^{2} = \epsilon_{p_{\lambda}}L/U_{0_{1D}}N$.

Moreover, the function $f(z)$ is given by
\begin{equation}
f(z)=1+z^2-z\sqrt{2+z^2}-\frac{1}{2z^2}.
\end{equation}
Notice that we have taken into account that the lower limit in the integral is not zero. This limit, that takes into account the energy of the ground state, and consequently, finite size effects of the system together with the polymer length contributions, is given by
\begin{eqnarray}
\gamma^{2} =\Bigl(\frac{L}{U_{0_{1D}}N}\Bigr)\frac{(\hbar/L)^{2}}{2m}\Bigl(1-\Bigl(\frac{\lambda}{L}\Bigr)^{2}\Bigr).
\label{Psedo33}
\end{eqnarray}
We must mention that if we take $\gamma=0$ (which corresponds to the case $p=0$) in (\ref{Psedo2}) the ground state energy  becomes divergent, and the condensation is never reached at finite temperature in the thermodynamic limit, within this approximation. We can see, then, that finite size effects of the system must be included in order to reach the condensation.

Finally, let us made some comments about the functional form of $U_{0_{1D}}$. We assumed that the one dimensional parameter $U_{0_{1D}}$ is given by \cite{tesis}
\begin{equation}
U_{0_{1D}}=-\frac{2\hbar^{2}}{ma_{1D}},
\end{equation}
where $a_{1D}$ is the scattering length in one dimension. This expression suggests that the one--dimensional scattering length is a function of the three dimensional one, $a_{3D}$, as in trapped Bose--Einstein condensates. Using these facts $a_{1D}$ seems to be
\begin{equation}
\label{a1d}
a_{1D}=-\frac{L^{2}}{2 a_{3D}}\Bigl(1-C\frac{a_{3D}}{L}\Bigr)\approx-\frac{L^{2}}{2 a_{3D}} .
\end{equation}
In the last term we have assumed that $a_{3D}$ is much smaller than $L$. In this approximation the one dimensional self--interaction parameter is thus given by $U_{0_{1D}}\approx 4\hbar^{2}a_{3D}/mL^{2}$.

\section{Speed of sound}

After performing the integrals in expression (\ref{Psedo2}), we are able to obtain the ground state energy $E_{0}$ associated with our \emph{polymer} N--body Hamiltonian 
\begin{eqnarray}
E_{0}&\approx& \frac{U_{0_{1D}}N^2}{2L} + \frac{(\hbar/L)^{2}}{2m}\Bigl[1-\Bigl(\frac{\lambda}{L}\Bigr)^{2}\Bigr]\,N \nonumber\\ &-&
\frac{L\sqrt{2m}}{4\pi\hbar}\Bigl(\frac{U_{0_{1D}}N}{L}\Bigr)^{3/2} \Bigl(\frac{2}{3}\sqrt{2}-\frac{1}{2 \gamma}-\gamma\Bigr)\nonumber\\&+&
3\lambda^{2} \frac{L\sqrt{2m^{3}}}{8\pi\hbar^{3}}\Bigl(\frac{U_{0_{1D}}N}{L}\Bigr)^{5/2}\Bigl(-\frac{18}{5}\sqrt{2}+\frac{1}{2}\gamma \Bigr).\,\,\,
\label{Psedo3}
\end{eqnarray}
It is noteworthy to mention that if we set the polymer length $\lambda=0$  we recover the \emph{usual} ground state energy obtained in Refs.\,\cite{paper60,paper60_1}, corrected by finite size contributions of the system, as in Ref.\,\cite{sp1}.

The corresponding squared speed of sound $c_{\lambda}^{2}$ is then given by
\begin{equation}
\label{SS}
c_{\lambda}^{2}=-\frac{L^{2}}{Nm} \Bigl(\frac{\partial P_{0}}{\partial L}\Bigr)=c_{1}+c_{2}+\lambda^{2}c_{3},
\end{equation}
where $P_{0}=-(\partial E_{0}/\partial L )$ is the ground state pressure. 

In the expression for the speed of sound (\ref{SS}) we have defined 
\begin{eqnarray}
c_{1}& = & \frac{24 a_{3D} \hbar^2 N}{L^3 m^2}-\frac{42 a_{3D} \hbar \sqrt{\frac{a_{3D} \hbar^2 N}{L^3 m}}}{\pi  L^2 m^{3/2}}, \nonumber
\end{eqnarray}
\begin{eqnarray}
 \nonumber\\&&c_{2}  =  \frac{198 \sqrt{2} a_{3D}^{2} \hbar N \sqrt{\frac{a_{3D} \hbar^2 N}{L^3 m}}}{\pi  L^3 m^{3/2}}+\frac{35 \hbar \sqrt{\frac{a_{3D} \hbar^2 N}{L^3 m}}}{8 \sqrt{2} \pi  L m^{3/2} N}+\frac{3 \hbar^2}{L^2 m^2},\,\,\,\,\,\,\,\,\,  \nonumber 
\end{eqnarray}
\begin{eqnarray}
c_{3}& = & -\frac{99 \hbar \sqrt{\frac{a_{3D} \hbar^2 N}{L^3 m}}}{8 \sqrt{2} \pi  L^3 m^{3/2} N}+\frac{429 a_{3D} \hbar \sqrt{\frac{a_{3D} \hbar^2 N}{L^3 m}}}{8 \sqrt{2} \pi  L^4 \sqrt{m^3}}-\frac{10 \hbar^2}{L^4 m^2}\nonumber \\
&&+ \frac{a_{3D}^{2} \left(78 \left(5 \sqrt{2}-54\right) \hbar N \sqrt{\frac{a_{3D} \hbar^2 N}{L^3 m}}\right)}{\pi  L^5 m^{3/2}},
\label{ss}
\end{eqnarray}
where $c_{1}$ corresponds to the usual result obtained in Ref.\,\cite{paper60,paper60_1}, $c_{2}$ are the finite size contributions as in Ref.\,\cite{sp1} and $c_{3}$ are the modifications associated with the polymer length scale $\lambda$. In other words, in the usual case $\lambda=0$, we recover the results obtained in Refs.\,\cite{paper60,paper60_1}  plus corrections due to the finite size effects of the condensate \cite{sp1}. 

In order to analyze the sensitivity of our system to  corrections caused by the polymer quantization we use the usual laboratory conditions: $N\sim 10^{4}$ particles, $a_{3D} \sim10^{-9}$ m, and $m\sim 10^{-26}$ kg,  \cite{Dalfovo} and $L\sim 10^{-3}$m. The speed of sound in a three--dimensional Bose--Einstein condensate is typically of order $10^{-3}$m$s^{-1}$ \cite{Andrews,Andrews2}. It is clear that although, from the experimental point of view there is no a real one--dimensional condensate, it is possible to construct a quasi--one--dimensional condensate just by using extremely anisotropic traps \cite{aniso}. Then, in principle, the use of this experimental accuracy is justified in order to analyze the sensitivity of the system to polymer corrections.

The experimental parameters mentioned above allows us to estimate the relative shift caused by the polymer length $\lambda$, which can be estimated from  
\begin{equation}
\frac{\Delta c_{\lambda}}{c_{\lambda=0}} \equiv \frac{c_{\lambda}-c_{\lambda=0}}{c_{\lambda=0}}\approx  -1.5\times10^{6} \,\lambda^{2}.
\end{equation}
If we cancel the polymer corrections, $\lambda=0$, we recover the usual value $c_{\lambda=0}$.

Taking into account the bound $\lambda^{2}\sim 10^{-16}$\,m, obtained in a previous work \cite{eli}, the relative shift in the speed of sound is of order $\sim 10^{-10}$\,m$s^{-1}$, at least seven orders of magnitude smaller than the typical value $10^{-3}\,$m$s^{-1}$ for $\lambda=0$, under typical experimental conditions.

However, due to the functional form of equation (\ref{ss}), densities of order $\frac{N}{L}\sim 10^{25}$ particles per meter, together with small values of the scattering length $a_{3D}\sim10^{-18}$ leads to a relative shift of up to $10^{-4}$\,m$s^{-1}$, assuming $\lambda\sim10^{-16}$.  In other words, dense systems together with small values of the scattering length $a_{3D}$ are needed in order to analyze the sensitivity caused by polymer quantization in one--dimensional condensates. 
The case for small values of $a_{3D}$ can be solved, in principle, just by tuning the interaction coupling by Feshbach resonances to very small values of the scattering length $a_{3D}$, that is, almost to the ideal case, and then reducing the contribution of interactions on the speed of sound below the \emph{polymer quantization} induced shift for dense enough systems. However, these assumptions could affect the stability of the condensate and could give rise to  technical difficulties. Furthermore, this scenario must be extended to a more realistic situation, i.e., one--dimensional trapped Bose--Einstein condensates in the lines of Refs.\,\cite{uwe}.

%\textbf{aqui analisis de las correcciones con datos experimentales}

%\textbf{Some comments on Polymer quantization in Trapped Bose-Einstein Condensates???}

%\textbf{ Propiedades para un condensado en trampa con operadores de creacion y aniquilacion corregidos con $\lambda$, referencias de esto...???}

\section{Conclusions}

Assuming that the Bogoliubov's formalism is valid, we have analyzed some properties associated with a one--dimensional Bose--Einstein condensate within the \emph{polymer quantization}. 
We obtained corrections, caused by the inclusion of the polymer length scale $\lambda$, on the ground state energy of this system, and consequently corrections to the corresponding speed of sound.
Moreover, we showed that high densities and small values of the interaction parameter are required to enhance the sensibility of our system to possible corrections caused by $\lambda$.  
The results obtained in this work must be extended to more realistic situations from the experimental point of view, namely, there are no condensates in a box. Usually the confinement of the condensate can be obtained by using harmonic traps, among others techniques \cite{yuka1,Dalfovo}, and the use of generic potentials, under certain conditions, could be useful in this context, because they change the shape of the trap \cite{CastellanosClaus,Castellanos}. In other words, the analysis of systems trapped in generic potentials within the Bogolioubov's formalism deserves further investigation, and will be presented in a forthcoming work \cite{elias2}. 
%\textbf{extender...}

Finally, we want to stress that the many-body contributions in a Bose--Einstein condensate open the possibility to explore specific scenarios that could be used, in principle, to analyze the sensitivity of these systems to effects of the discreteness of the space.

\begin{acknowledgments}
E.C.  acknowledges UNACH/MCTP for financial support. H. H. H. also acknowledges the grant PRODEP UACH-CA-125.
\end{acknowledgments}


\begin{thebibliography}{}


\bibitem{eli} E. Castellanos and G. Chac\'{o}n-Acosta, Phys. Lett. B \textbf{%
\ 722,} 119 (2013).

\bibitem{CastellanosCamacho} E.~Castellanos and A.~Camacho, Gen. Rel. Grav.
\textbf{41}, 2677 (2009).

\bibitem{CastellanosCamacho1} E.~Castellanos and A.~Camacho, Mod. Phys.
Lett. A \textbf{25}, 459 (2010).

\bibitem{Castellanos} E. Castellanos and C. Laemmerzahl, Mod. Phys. Lett. A
\textbf{27}, 1250181 (2012).

\bibitem{r1} F. Briscese, M. Grether, and M. de Llano, Europhys. Lett.
\textbf{98}, 60001 (2012).

\bibitem{r2} F. Briscese, Phys. Lett. B \textbf{718}, 214 (2012).

\bibitem{elca} A. Camacho and E. Castellanos, Mod. Phys. Lett. A, \textbf{27}%
, 1250198 (2012).

\bibitem{eli2} E. Castellanos, Europhys. Lett. \textbf{103}, 40004 (2013).

\bibitem{CastellanosClaus} E. Castellanos and C. Laemmerzahl, Phys. Lett. B
\textbf{731}, 1 (2014).

\bibitem{ECA}
E. Castellanos, J. I. Rivas and V. Dominguez-Rocha, Europhys. Lett, \textbf{106},  60005 (2014)

\bibitem{ECA1}
E. Castellanos, J. I. Rivas, Phys. Rev. D \textbf{91}, 084019 (2015).

\bibitem{ECe}
El\'ias Castellanos, Celia Escamilla-Rivera, arXiv:1507.00331 (2015). 

\bibitem{AS1}
A. Ashtekar, S. Fairhurst and J. L. Willis, Class. Quantum Grav. \textbf{20}, 1031 (2003).


\bibitem{AS2} 
A. Corichi, T. Vukasinac and J. A. Zapata, Phys. Rev. D \textbf{76}, 044016 (2007); A. Corichi, T. Vukasinac and J. A. Zapata, 
Class. Quantum Grav. \textbf{24}, 1495 (2007).

\bibitem{Gam}
R. Gambini and J. Pullin, Phys. Rev. D {\bf59}, 124021 (1999). 

\bibitem{hugo}
J. Alfaro, H. A. Morales-Tecotl and L. F. Ueeutia, Phys. Rev. Lett. {\bf84}, 2318 (2000). 

\bibitem{gio1}
G. Amelino-Camelia and S. Majid, hep-th/9907110, Int. J. Mod. Phys. A {\bf15} 4301 (2000).

\bibitem{j}
J. Kowalski-Glikman, astro-ph/0006250, Phys. Lett. B {\bf 499} 1 (2001).

\bibitem{O}
O. Bertolami and L. Guisado, hep-th/0306176, JHEP 0312, 013 (2003).

\bibitem{sr}
S. R. Coleman and S. L. Glashow, Phys. Rev. D {\bf59}, 116008
(1999).

\bibitem{chaca}
Guillermo Chacon-Acosta, Hector Hernandez-Hernandez, Int. J. Mod. Phys. D \textbf{24}, 1550033 (2015).

\bibitem{vel} 
J. M. Velhinho, Class. Quantum Grav. \textbf{24}, 3745 (2007).

\bibitem{viqar}  V. Husain and A. Kreienbuehl, Phys. Rev. D {\bf 81}, 084043 (2010); M. Arzano, Phys. Rev. D {\bf 90}, 104036 (2014).

\bibitem{Karabulut}
 E. Karabulut, M. Koyuncu, M. Tomak, Physica A \textbf{389}, 1371 (2010).

\bibitem{yan} 
Z. Yan, Phys. Rev. A {\bf 59}, 4657 (1999).

\bibitem{Dalfovo}
F. Dalfovo, S. Giordini, L. Pitaevskii and S. Strangari, Rev. Mod. Phys. 71, 463 (1999).

\bibitem{BP}
B. Paredes et al., Nature (London) \textbf{429}, 277 (2004).

\bibitem{TK}
T. Kinoshita, T. Wenger, and D. S. Weiss, Science \textbf{305}, 1125 (2004) .

\bibitem{sp1}
El\'ias Castellanos, \emph{Homogeneous one-dimensional Bose-Einstein Condensate in the Bogoliubov's Regime}, arXiv:1604.01730v2 [cond-mat.quant-gas] (2016).

\bibitem{chiapas} 
G. Chac\'{o}n-Acosta, L. Dagdug and H. Morales-T\'ecotl, AIP Conf. Proc. \textbf{1396}, 99 (2011).

\bibitem{yuka1}
V. I. Yukalov, Phys. Rev. A {\bf 72}, 033608 (2005).

\bibitem{paper60}
Elliott H. Lieb and Werner Liniger, Phys. Rev. \textbf{130}, 1605 (1963).

\bibitem{paper60_1}
Elliott H. Lieb, Phys. Rev. \textbf{130}, 1616 (1963) .

\bibitem{tonks}
L. Tonks, Phys. Rev. \textbf{50}, 955 (1936); M. Girardeau, J. Math. Phys. \textbf{1}, 516 (1960); A. Lenard, J. Math. Phys.,
\textbf{7}, 1268 (1966).

\bibitem{FB}
T. Cheon and T. Shigehara, Phys. Rev. Lett. 82, 2536 (1999); V. E. Korepin, N. M. Bogoliubov, and A. G. Izergin, Quantum Inverse Scattering Method and Correlation Functions (Cambridge University Press, Cambridge, 1993).

\bibitem{cigarro}
V. Dunjko, V. Lorent, and M. Olshanii, Phys. Rev. Lett. \textbf{86} (24), 5413-6 (2001).

\bibitem{Bog}
N. N. Bogoliubov, J. Phys (U.S.S.R.) \textbf{11}, 23 (1947)

\bibitem{Bog1}
N. N. Bogoliubov and D. N. Zubarev, Sov. Phys.-JETP \textbf{1}, 83 (1955).

\bibitem{Ueda}
M. Ueda, \emph{Fundamentals and New Frontiers of Bose--Einstein Condensation}, World Scientific, Singapore,
(2010).

\bibitem{pathria} 
R. K. Pathria, Statistical Mechanics (Butterworth, 1996).

\bibitem{tesis}
Igor A. Romanovsky, \emph{Ph.D., tesis}, School of Physics Georgia Institute of Technology (2006).

\bibitem{Andrews}
M. R.~Andrews et al, Phys. Rev. Lett. \textbf{79}, 553--556 (1997).

\bibitem{Andrews2}
M. R.~Andrews et al, Phys. Rev. Lett. \textbf{80}, 2967--556 (1998).

\bibitem{aniso}
A. Gorlitz, et al., Phys. Rev. Lett. \textbf{87} 130402 (2001) .

\bibitem{uwe}
Uwe R. Fischer, Phys. Rev. Lett. \textbf{89}, 280402 (2002); Uwe R. Fischer, Journal of Low Temperature Physics
\textbf{138} (3), 723-728 (2005).

\bibitem{elias2}
E. Castellanos, G. Chacon-Acosta, \emph{Polymer Quantization in One--Dimensional Trapped Bose-Einstein Condensates} (work in
progress).


\end{thebibliography}
\end{document}